\documentclass[11pt,twoside]{article}

\usepackage{asp2004}
\usepackage{epsf}
\usepackage{psfig}
\usepackage{lscape}

\newcommand{\msun}{\ensuremath{\mathit{M}_{\odot}}}                  

\begin{document}
\title{Observational Constraints on Cluster Evolution}    
\author{S{\o}ren S. Larsen}   
\affil{Astronomical Institute, University of Utrecht, The Netherlands}    

\begin{abstract} 
  Current observational constraints on the dynamical evolution of star
clusters are reviewed. Theory and observations now agree nicely on
the mass dependency and time scales for disruption of young star clusters
in galactic disks, but many problems still await resolution.  The origin 
of the mass function of old globular clusters, and its (near) invariance 
with respect to host galaxy properties and location within the host galaxy 
remain prominent puzzles.  
Most current models fail to reproduce the globular cluster mass function as a 
result of dynamical evolution from an initial power-law, except under very
specific conditions which are not generally consistent with observations.
How well do we actually know the proper initial conditions?
The cluster initial mass function (CIMF) 
seems to be consistent with a power-law with exponent $\alpha\approx-2$ 
in most present-day star forming galaxies, but the limits of the mass range
over which this approximation is valid
remain poorly constrained both observationally and 
theoretically. Furthermore, there are hints that some dwarf galaxies may 
have CIMFs which deviate from a power-law.  
\end{abstract}


\section{A bit of historical background}

  The idea that star clusters evolve dynamically has been around for nearly
as long as the very concept of star clusters itself. The first to discuss
the general properties of clusters in some detail was William Herschel, who 
attributed their various degrees of ``central concentration'' to an 
evolutionary sequence in which the globular clusters represented the final
stage \citep{wh1789}. At that time, however, a real theory of stellar dynamics 
was still far in the future and the subject of cluster evolution remained 
dormant for well over a century.

In the early 20th century, the attention was mostly focussed on external 
perturbations.  \cite{jeans1916} remarked that ``A cluster of stars will 
suffer continual disintegration from its encounters with other stars or 
clusters''.  In his monograph on ``Star Clusters'', \citet[][p.\ 208]{shap1930} 
devoted only a brief statement to dynamical evolution, noting that ``some of 
the globular clusters may be affected in freedom, form, and eventual survival 
by contacts with galaxies or other clusters.'' The stability of ``moving 
clusters'' was investigated by Jeans and later by \citet{bok1934}.  The 
important effect of internal two-body encounters was first studied by 
\citet{amba1938} who noted that stars in the high-velocity tail of the 
Maxwellian velocity distribution will gradually escape, leading to disruption 
of typical open clusters on time scales of a few Gyrs, depending on cluster 
mass. With remarkable foresight, Ambartsumian also realised that low-mass stars 
will escape preferentially, and suggested that more evolved clusters might be 
recognisable by their lack of low-mass stars. Similar conclusions were reached 
(in the context of old globular clusters) by \citet{spitzer1940}.  These ideas 
have been continuously refined ever since and it would lead too far to review 
the relevant literature on models for the dynamical evolution of star clusters 
here. Excellent reviews on the dynamical evolution of open and globular 
clusters are in \cite{king1980}, \cite{spitzer1987} and \cite{mh1997} and the 
field remains very active as illustrated by several contributions in this 
volume.  The remainder of this review concentrates on observational 
constraints on cluster evolution.

In order to address dynamical evolution in a meaningful way observationally, 
a method to age-date individual star clusters becomes a necessity.  The 
catalogue of open clusters compiled by \citet{trump1930} listed the spectral 
types of the brightest stars in 100 Milky Way open clusters, but the use of 
this information as a suitable chronometer had to await the emergence of the 
first detailed sets of stellar model calculations in the 1950s.  Once it 
became possible to derive cluster ages from colour-magnitude diagrams, it 
quickly became clear that the number of old open clusters is lower than one 
would naively predict from a continuous formation rate, even taken into account 
the fact that younger clusters are more easily detected than older ones because 
they contain brighter stars.

The relative paucity of old open clusters was noted almost simultaneously 
in at least three papers \citep{vdb1957,oort1958,vonh1958}, all of which based 
on Trumpler's catalogue.  It appears that they all arrived at the same 
conclusion quite independently (van den Bergh, priv.\ comm.).  All three papers 
also singled out disruption as the most likely cause of the decline in number 
of clusters with ages greater than $\sim10^8-10^9$ years.  Considering the 
effects of stellar mass loss, a tidal field and relaxation due to stellar 
encounters, \citet{vonh1958} estimated an average cluster lifetime of $10^9$ 
years, in reasonable agreement with his best observational estimate of 500 
Myrs.  \citet{spitzer1958} calculated cluster disruption times of $10^8-10^9$ 
years for typical open cluster densities, in fair agreement with the 
observational estimates, and also noted that encounters between clusters and 
interstellar clouds might play an important role for cluster disruption.  
Thus, one might argue that theoretical and observational estimates of cluster 
disruption times were in agreement to within a factor of 2 already 50 years 
ago! In the context of this meeting, it is also interesting to note that the 
connection between stellar mass loss and the evolution of star clusters was 
recognised early on. Of course, many of the details still remained (and 
remain!) to be worked out. 

Using more recent compilations of open cluster data, \citet{wielen1971}
estimated that about 50\% of open clusters in the Milky Way disintegrate
within about 200 Myrs. He also noted a significant scatter in the disruption 
times, with 2\% of the clusters surviving for longer than a Gyr, and suggested 
that the disruption time probably depends on the size and mass of the cluster 
(as anticipated by Ambartsumian).  Observational support for the importance of 
giant molecular cloud (GMC) encounters was provided by \citet{vm1980} who 
pointed out that the oldest open clusters ($>1$ Gyr) are strongly concentrated 
towards the galactic anti-centre direction, where the density of GMCs is lower. 
This observation, as well as the fact that old open clusters tend to be found 
further from the Galactic plane than young ones, has since been confirmed by 
several studies \citep[][and references therein]{friel1995}. Theoretically, 
further support for the importance of GMC encounters came from N-body 
simulations by \citet{terle1987}.

The first observational evidence that cluster disruption times depend on mass 
came from \citet{ja1982}.  However, it has also been clear for some time that 
the disruption times may not be the same in all galaxies.  Both the LMC and 
SMC show a less pronounced lack of old clusters than the Milky Way 
\citep{ef1985,hodge1987}, and already \citet{wielen1985} suggested that 
different GMC densities in the Milky Way, SMC and LMC might play a role.  
However, a direct comparison with the Milky Way remained somewhat hampered by 
different detection limits and the bursty star formation history of the LMC. 
Prior to the launch of the \emph{Hubble Space Telescope} in 1990, little was 
known about the detailed properties of (young) cluster populations in other 
galaxies, and even in the Local Group spirals M33 and M31 the data were 
insufficient to strongly constrain cluster lifetimes.  

Little has been said explicitly thus far about the \emph{globular} clusters 
(GCs), which present a number of interesting problems.  In the Milky Way, they 
are associated with the spheroidal component(s) and thus considered typical 
``Population II'' objects \citep{baade1944}. In fact, every major galaxy is 
surrounded by ancient GCs akin to those found in the Milky Way.  The mass 
distribution of GCs is markedly different from that of Milky Way open clusters, 
with a relative deficit of low-mass GCs \citep[\S\ref{sec:gclf},][]{vl1984}. 
This difference has sometimes lead to the notion that GCs may be fundamentally 
different from star clusters forming today, and might have formed by quite 
different processes which were unique in the early Universe \citep{pd1968}. 
However, the young ``populous'' star clusters in the LMC, some of which have 
GC-like masses, have remained a puzzle since the early 20th century 
\citep{shap1930}, and 
continue to serve as a reminder that the distinction between open and 
globular clusters is not necessarily as clear-cut in other galaxies as it 
may seem in our own.  

The idea that GC formation may not have been restricted to the early Universe 
received renewed interest when \citet{schweizer1987} suggested that new GCs 
form in galaxy mergers. This might solve the ``specific frequency problem'' 
(i.e., the fact that elliptical galaxies contain many more GCs per unit host 
galaxy luminosity than do spirals), which had been raised as a major objection 
against the idea that elliptical galaxies form from spiral-spiral mergers 
\citep{vdb1982} as suggested by \citet{tt1972}. The subsequent discovery by 
HST observations of young ``super'' star clusters in a number of interacting 
and merging galaxies in the 1990s \citep[as reviewed by][]{whitmore2003} 
further stimulated interest in this idea, but the problem remains that the
``cluster initial mass function'' (the CIMF) seems very different from that of 
the ancient GCs. Besides other objections to the major merger picture 
\citep*[e.g.][]{fbg1997} this still represents a major challenge, as I will 
discuss shortly.

\section{Evolution of cluster populations in star-forming galaxies}

  Over the last 1--2 decades, it has become abundantly clear that the basic 
division into ``open'' and ``globular'' clusters as it applies in the Milky Way 
quickly becomes inadequate when a wider variety of host galaxy types are 
considered. Young clusters with masses in the range $10^4-10^6$ \msun\ or 
greater have now been found in many different types of external galaxies 
\citep{larsen2005}.  Furthermore, HST observations of some lenticular (S0) 
galaxies have revealed populations of old ($\ga8$ Gyrs) star clusters with 
larger effective radii than normal open and globular clusters, and with a mass 
distribution that does not display the turn-over observed in classical old GC
systems \citep{lb2000,bl2002,peng2006}. These ``faint fuzzy'' clusters are 
clearly associated with the disks of their host galaxies, and do not easily 
fit the classical description of either open or globular clusters.

  Ultimately, the dynamical evolution of all flavours of star clusters must 
be governed by the same basic laws of stellar dynamics. We 
must therefore demand from any successful theory of cluster evolution that 
it is able to account for the observed properties of \emph{all} types 
of star clusters without too much tweaking. The fact that the luminosity
function of old globular clusters is (nearly) universal may be an important
clue to the mechanisms that shaped it, as will be discussed further
below (\S\ref{sec:gclf}). However, before seeking
to explain the \emph{evolution} of clusters, we must make
sure that we know their \emph{initial} properties. 

\subsection{The Cluster (Initial) Mass Function}

Was the difference between the MFs of young and old (``globular'')
star clusters set up at formation, or is it a result of dynamical evolution? 
That GCs formed with a MF that was different from that observed in young 
cluster systems today \citep{pg2005} is perhaps not an entirely unreasonable 
idea, given that conditions in the early Universe were likely quite 
different from those prevalent
today. However, this is unattractive for several reasons: First, 
direct observations of the formation of individual GCs at cosmological 
distances to test this hypothesis (over a mass range that allows useful 
constraints on the CIMF) will not be possible for a long time to 
come.  Second, it would be
a return to the notion that GCs are ``special'', and thus make them less
attractive as general tracers of the star formation histories of galaxies.
At the end it may of course turn out that this is how Nature really works.
But alternatively, we may seek to understand the present-day MF
of GCs as a result of dynamical evolution from the MFs observed in
young cluster systems. It is then worthwhile to ask whether there \emph{is}
a universal CIMF, and what it might look like.

  Even after more than a decade of HST observations, the number of galaxies
where the MF of young star clusters is known well is still in the single 
digits.  Direct application of the virial theorem requires reliable 
measurements of both the size and velocity dispersion of each individual 
cluster, and is impractical for large samples of clusters.
Most determinations of the masses of extragalactic 
star clusters rely on assumptions about the mass-to-light (M/L) ratio, usually 
from simple stellar population (SSP) models which assume a ``standard'' 
stellar mass function \citep[e.g.][]{salpeter1955,kroupa2001}. The major 
complication here is that the M/L ratios are strongly age-dependent, so that 
knowledge of individual cluster ages is required in order to convert the 
observed luminosities into a MF.  SSP models provide predictions 
for the evolution of broad-band colours as a function of age, which can 
be used to infer the ages of star clusters for (in principle) arbitrary 
combinations of passbands.  However, for clusters younger than about 
a Gyr, optical colours are only weakly sensitive to age, and strongly
degenerate with respect to reddening, making imaging in the $U$-band a
necessity for reliable age (and hence mass) determinations.  

\begin{figure}[!ht]
\plotone{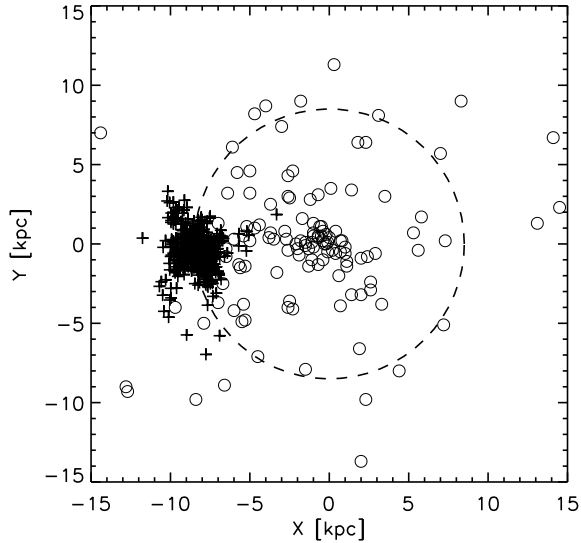}
\caption{\label{fig:mwclusters}The spatial distribution of catalogued
open clusters (plus markers) and globular clusters (open circles) in 
the Milky Way. The Sun is at (X,Y) = (-8.5, 0) kpc.}
\end{figure}

  Galaxies where the MF has been determined for young clusters include
the Milky Way \citep{ee1997}, the LMC and SMC \citep{hunter2003}, 
the nearby interacting spiral M51 \citep{bik2003},
the starburst galaxies NGC~3310 and NGC~6745 \citep{deg2003a} and
the ``Antennae'' \citep{zf1999}. In all these cases, the MF
is consistent with a power-law of the form
$dN/dM \propto M^\alpha$ with $\alpha\approx-2$, although the mass
ranges probed by the different studies differ substantially. In the
Antennae and in NGC~3310/NGC~6745, the lower limits are at 
$>10^4$ \msun, but these are purely observational limits and there
is no reason to suspect that lower-mass, ``open'' clusters are absent 
in these galaxies. This lower limit coincides roughly with the
mass of the most massive young clusters known in the Milky Way
(with a couple of exceptions), but
again this may be an observational selection effect due to the limited
volume of the Galactic disk probed by current open cluster catalogues.

  Fig.~\ref{fig:mwclusters} shows the spatial distribution of open and globular 
clusters in the Milky Way from the catalogues of \citet{khar2005} and 
\citet{harris1996}, projected onto the Galactic plane. The median distance of 
the open clusters from the Sun is 1.1 kpc, and more than 80\% have distances 
$<2$ kpc, a strong indication that the current sample is highly spatially 
incomplete beyond $\sim1$ kpc.  Indeed, a few young clusters with masses of 
$\sim10^5$ \msun\ have recently been identified in the Milky Way, including 
Westerlund 1 \citep{clark2005} and an object detected in the 2MASS survey 
\citep{figer2006}. Both are located at distances of several kpc (but still 
closer than the Galactic centre), and subject to large amounts of foreground
extinction. The existence of these objects at distances of a few kpc is 
fully consistent with extrapolation from the MF of open clusters in the Solar 
neighbourhood, and strongly suggests that several young clusters with masses 
of $10^4-10^5$ \msun\ or greater remain undiscovered in the disk of our Galaxy 
\citep{larsen2006}.  The known globular clusters, on the other hand, are more 
or less symmetrically distributed with respect to the Galactic centre,
suggesting that the sample does not suffer from severe spatial incompleteness.

\subsection{The Poor Man's Approach: Luminosity Functions}

Luminosity functions (LFs) are more readily obtained for large samples of 
clusters than MFs.
Generally, these also show power-law behaviours although often with
slightly steeper slopes than the MFs \citep{whit1999,larsen2002}. One way 
such a difference could arise is if the MF is truncated at 
some upper limit $M_{\rm max}$. \citet{whit1999} noted a ``bend'' in the
LF of young clusters in the Antennae at $M_V\sim-10.4$,
with a slope of $\alpha=-2.6$ above the bend. For an age of 10 Myr, 
the bend corresponds to a mass of $\sim10^5$ \msun, tantalisingly close to
the turn-over of the globular cluster MF.
However, \cite{zf1999} found that 
the MF of young clusters in the Antennae is, in fact, consistent 
with a power-law with $\alpha=-2$ over the range $10^4-10^6$ \msun, and
noted that the bend in the LF could be due to a truncation of the MF near
$10^6$ \msun .

The relation between the presence of a bend in the LF
and a truncation of the MF has been investigated in more
detail by \citet{gieles2006a,gieles2006b}, who suggested that the
MF in the spirals M51 and NGC~6946 is truncated
at $M_{\rm max} =0.5-1\times10^6$ \msun, and at about $2\times10^6$ \msun\
in the Antennae. Thus, while direct data for the MF are preferable,
studies of the LF may still hold interesting
clues to the underlying MF, albeit not without some assumptions.

It must be mentioned that some dwarf galaxies host a few star
clusters which appear too luminous (and massive) to simply form the
high-mass tail of a power-law distribution. Most conspicuous among
these is NGC~1705, in which there is a gap of 3 magnitudes between
the two brightest clusters, but a similar effect is observed in other
dwarfs \citep*{bhe2002,johnson2005}. In an HST study of a sample including 
36 dIrr
galaxies, \citet*{shar2005} noted a possible turn-over in the LF
of young star clusters, and speculated that different cluster
formation mechanisms might be at work in isolated dwarf galaxies.
Such differences would be particularly interesting in view of the idea 
that some GCs may have formed in dwarf-like fragments \citep{sz1978,pg2006}.
Inevitably, studies of star clusters in dwarf
galaxies tend to be complicated by small number statistics, and
a systematic study of an even larger sample of star-forming dwarfs would be
desirable. 

In summary, the current (limited) evidence suggests that the CIMF is 
consistent with a 
power-law $dN/dM\approx M^{-2}$ \emph{over some mass range} in 
most galaxies.  So far, there is no indication for any significant difference 
in the shape of the CIMF in the Milky Way and more actively star-forming 
galaxies. If an upper limit to the CIMF exists in our Galaxy it is
probably at $10^5$ \msun\ or greater,
and the apparent lack of ``super star clusters'' in the 
Milky Way disk may be, at least to some extent, a size-of-sample effect 
due to the limited volume we are sampling. More generally, whether or not
the CIMF has an upper limit, and how it might depend on host
galaxy properties, is only now starting to be explored.
However, we must also remain open to the possibility that cluster populations 
do not \emph{everywhere} form with a single power-law MF.

\subsection{Cluster disruption in disks}

Although the basic mechanisms responsible for the disruption of star clusters 
were identified already half a century ago, a detailed framework to compare 
observations and theory has only recently become available. 

\citet{bl2003} derived the ``disruption time'', $t_{\rm dis}$, for 
star clusters in M51, M33, the SMC and the Solar neighbourhood under 
the assumption that $t_{\rm dis}$ could be parameterised as a simple 
analytic function of cluster mass: $t_{\rm dis} = t_4 (M/10^4\msun)^\gamma$,
where $t_4$ is the disruption time of a $10^4$ \msun\ cluster. By
explicitly including the dependence on mass, they could
compare disruption times in different galaxies more directly than
had been done before. The longer cluster disruption time in the SMC 
(by about an order of magnitude) was confirmed, and \citet{bl2003}
also found a significantly shorter disruption time in the
central parts of M51. Interestingly, the parameter $\gamma$ was found to 
have a value close to 0.60. This is significantly shallower
than expected from classical two-body relaxation, which predicts that a
star cluster should lose mass at a roughly constant rate and thus
$\gamma\approx1$ \citep{spitzer1987,fz2001}. However, a value of
$\gamma\approx0.62$ is consistent with $N$-body simulations for
clusters with a realistic stellar mass function (albeit with concentrations 
typical of GCs), evolving in an external tidal field 
\citep*{bm2003,lamers2005a}. The scaling factor $t_4$ appears to depend
on the mean ambient density as expected from $N$-body simulations, although 
the disruption time derived for the central regions of M51 by 
\citet{lamers2005a} remained too short by about an order of magnitude
compared to the predictions.
The shorter disruption time-scale near the centre of M51 might be caused
by a high density of GMCs there \citep{gieles2006c}.

\begin{figure}
\plotone{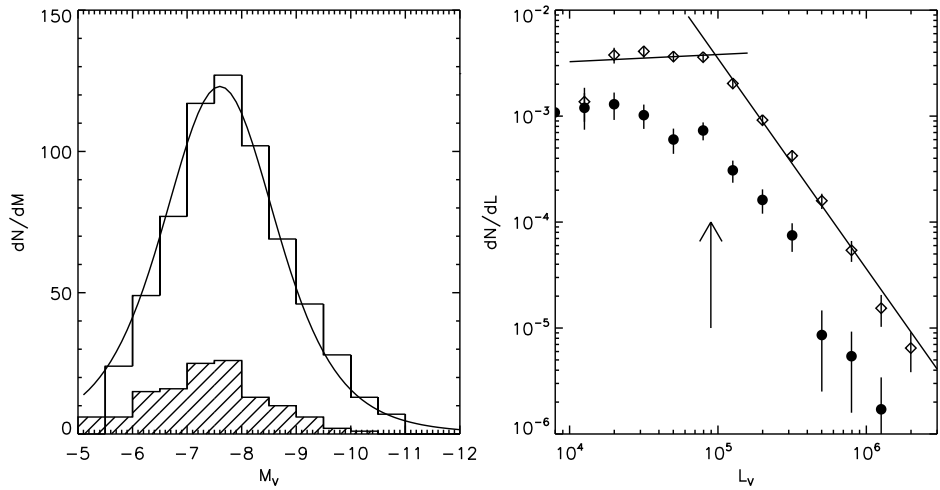}
\caption{\label{fig:gclf}The luminosity functions for GCs in the Sombrero 
galaxy \citep[data from][]{spitler2006} and the Milky Way
($R_{gc} < 80$ kpc). Left: number of GCs per magnitude bin and the best 
fitting Student's $t_5$-function. The Milky Way GCLF is shown as a hashed
histogram.  Right: Same data, but shown as number of GCs per luminosity bin. 
Milky Way data are shown with filled circles.  The arrow marks the luminosity 
corresponding to the peak of the $t_5$ fit. The straight lines are power-law
fits to the Sombrero data below and above the peak of the $t_5$ fit.}
\end{figure}

\citet{bl2003} assumed that clusters disrupt instantaneously at $t_{\rm dis}$, 
but in later papers a more realistic treatment of gradual mass loss has been 
included.  \citet{lamers2005b} showed that the present-day age- and mass
distributions of Milky Way open clusters can both be very well fitted in a 
scenario where clusters form at a (nearly) constant rate and disrupt on a 
timescale proportional to $M^{0.62}$. They derived a disruption time for a 
$10^4$ \msun\ cluster in the Solar neighbourhood of 1.3 Gyr, based on the 
\citet{khar2005} catalogue. For masses in the range $10^2 < M/\msun < 10^3$, 
this corresponds to disruption times between 75 and 300 Myrs, consistent with 
the earlier studies.

A few words on the concept of a high degree of ``infant mortality'' for
star clusters, which has become popular in recent years: This idea
is inspired by the facts
that 1) the formation rate of embedded star clusters in the Milky Way
over-predicts the number of observed open clusters by a large factor
\citep{ll2003} and 2) many studies of extragalactic young
star clusters find a disproportionately large number of objects
with ages $<10^7$ years \citep*{fall2005,bastian2005}.  The
objection has been raised that these young objects may not be
gravitationally bound, and thus do not deserve the label ``star clusters'' 
to begin with \citep{schweizer2006}. Certainly, young
clusters tend to be located in crowded regions, and the 
identification of individual clusters becomes challenging even with
HST at distances greater than a few Mpc \citep{larsen2004}. Ultimately,
whether or not the definition of a star cluster should include the
requirement that it be gravitationally bound may depend on the context,
and for very young, still embedded objects it may be impossible to
apply this criterion. In any case, this initial round of destruction 
appears to be independent of mass, and should not strongly affect the 
shape of the MF. 

\section{The Globular Cluster Mass Function}
\label{sec:gclf}

One of the more remarkable and puzzling facts about old GCs is the (near) 
universality of their luminosity function -- the GCLF.  Assuming that GCs 
in most galaxies display at most a small age spread, the GCLF is a good proxy 
for the MF.  Fig.~\ref{fig:gclf} compares the Milky Way GCLF 
[data from \citet{harris1996}] with data for the Sombrero galaxy, based on 
HST/ACS observations \citep{spitler2006}.  In the left panel, the two GCLFs are 
plotted in the traditional way, i.e.\ as a histogram of number of clusters 
per magnitude bin. The best Student's $t_5$ function fit to the
Sombrero data from Spitler et al.\ is overplotted. Clearly, the GCLFs
in the two galaxies are very similar.
That the GCLF may have a universal shape
was suggested early on by \citet{hr1979}, and it is now clear that the
variations, if any, are small over a large range of Hubble type
and luminosity \citep[e.g.][]{richtler2003,strader2006}. It should be noted 
that 
there \emph{are} subtle radial trends in the GCLF within the Milky Way GC 
system. In particular, the luminosity function of GCs with galactocentric 
distances $>80$ kpc may be bimodal \citep{vdb2003}. These outer halo clusters 
have been omitted from Fig.~\ref{fig:gclf} although they are 
relatively few in numbers and would not strongly affect the comparison.

The impression of a characteristic GC
mass is driven to some extent by the logarithmic binning of the data
\citep{mclaughlin1994,richtler2003}. When the GCLF is plotted as number of 
clusters per luminosity
bin it can be equally well fitted by a broken power-law as
illustrated in the right-hand panel of Fig.~\ref{fig:gclf}. The arrow
marks the peak of the $t_5$ function fit and the two straight lines
are power-law fits to the Sombrero GCLF on either side of the arrow.
At the high-mass end, the slope is $dN/dL\propto L^{-1.98\pm0.08}$,
very close to the ``canonical'' $\alpha=-2$ value for the MF
in young cluster systems. Similar results have been found for many other
GC systems \citep{hp1994,larsen2001}, although the exact slope evidently
depends on the mass range over which the fit is carried out.
The detailed behaviour of the GCLF below the turn-over is less well
constrained, which is unfortunate since this is where the MFs of young and 
old cluster systems differ. It probably cannot yet be ruled out that some
variations are present.  In the case of the Sombrero, the
fit in Fig.~\ref{fig:gclf} indicates an essentially flat distribution,
but possibly slightly steeper for the Milky Way.

\subsection{Radial Trends (or lack thereof)}

\begin{figure}
\plotone{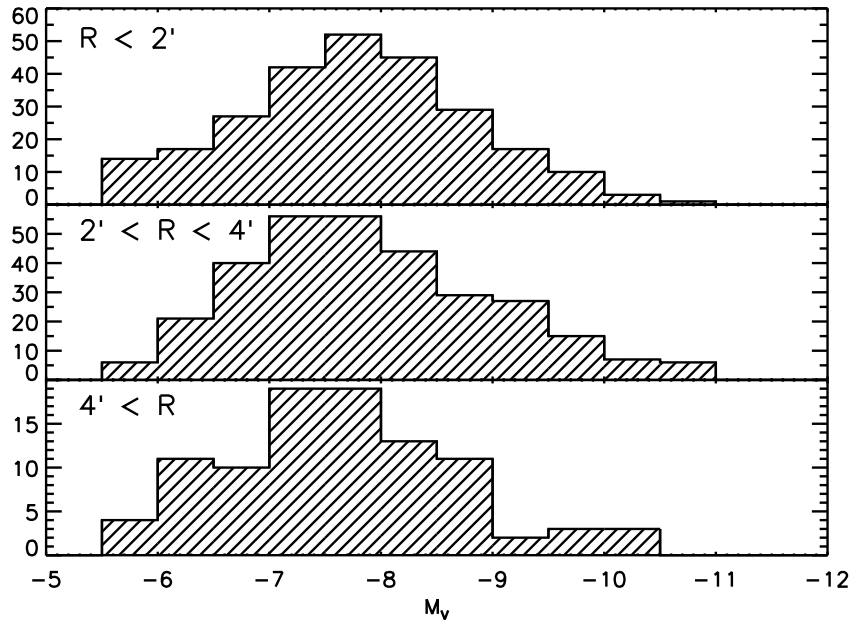}
\caption{\label{fig:sombrero}Luminosity function for GCs in the Sombrero 
galaxy in three radial bins. At a distance of 9.0 Mpc, the bins correspond 
to $0<R<5.2$ kpc, $5.2<R<10.4$ kpc and $R>10.4$ kpc. There is no 
statistically significant difference between the GCLFs in the three bins.}
\end{figure}

Fig.~\ref{fig:sombrero} shows the Sombrero GCLF in three radial bins,
corresponding roughly to (projected) galactocentric distances
of $0<R<5$ kpc, $5<R<10$ kpc and $R>10$ kpc.  It is clear that the GCLF
does not change much with radial distance, although a visual inspection of 
the histograms might suggest a slight shift of the peak towards 
fainter magnitudes in the
outer bins. However, differences in the turn-over are not 
statistically significant \citep{spitler2006}

Although it has long been clear that dynamical evolution will lead
to preferential survival of star clusters with certain combinations of
mass and size \citep[e.g.][]{fr1977}, this
invariance of the GCLF with galactocentric radius is a challenge in most 
models for the dynamical evolution of the MF. Qualitatively, most models 
develop a bend or turn-over as the low-mass clusters disrupt, but disruption 
should be more effective closer to the centre, invariably leading to a 
decrease in turn-over mass with galactocentric 
distance.  \citet{fz2001} found that the absence of a significant radial
trend in the Milky Way GCLF turn-over could be explained if the velocity 
distribution becomes more radially anisotropic with galactocentric distance.
However, the required degree
of radial anisotropy is inconsistent with the observed GC kinematics,
although the authors argued that an initial population of GCs on highly
radial orbits might have been disrupted by now.  In their model the 
cluster disruption time scales linearly with mass (i.e.\ $\gamma=1$), 
contrary to the empirical findings of \citet{bl2003} 
and \citet{gieles2005} and the $N$-body simulations of \citet{bm2003}. 
For the GC system of M87, \citet{vespe2003} again found that 
the observed constancy of the GCLF turn-over requires strongly radial orbits
beyond a few kpc, at odds with kinematic data \citep{cote2001}.

\begin{figure}
\plotone{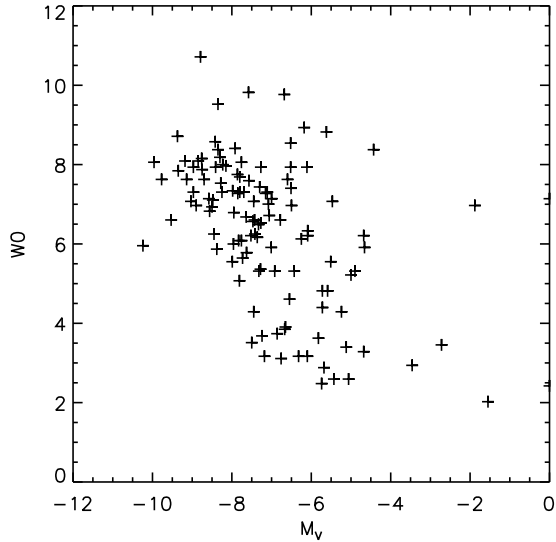}
\caption{\label{fig:mv_w0}Concentration parameter versus $M_V$ magnitude
for Milky Way globular clusters [data from \citet{harris1996}].
}
\end{figure}

One shortcoming of many models is the assumption of a single concentration
parameter. In fact, for GCs in the Milky Way the concentration parameter
displays a correlation with $M_V$ magnitude. Fig.~\ref{fig:mv_w0}
shows the \citet{king1966} concentration parameter $W_0$ versus $M_V$
for the Galactic GCs. While there is a significant scatter in
$W_0$ at any given $M_V$, there is also a clear trend of decreasing
concentration for fainter clusters. \citet{vz2003} found that
early stellar mass loss may lead to disruption of low-concentration 
clusters, causing a turn-over in the MF at about the right mass. Since 
this early disruption is mainly driven by processes internal to the cluster,
the lack of dependence on external parameters would be naturally explained.
However, a detailed study of this effect on the time evolution of the
MF is yet to be carried out.

\subsection{Departures from a universal GCLF?}
\label{sec:departures}

  Although the previous paragraphs have stressed the similarity of the
GCLF in different environments, there are some hints that the GCLF
may not be truly universal. As already noted, the GCLF in the outer
part of the Milky Way halo shows some differences with respect to
the canonical shape illustrated in Fig.~\ref{fig:gclf} and appears
skewed towards fainter luminosities. A similar effect may be present
in the sample of GCs in dwarf galaxies studied by \citet{shar2005}.
In a sample of five spiral galaxies, \citet*{chandar2004} noted an
excess of faint, red star clusters in M101 and, perhaps, NGC~6946.
In the case of M101 this excess was confirmed by \citet{barmby2006}, 
but these authors suggested that many of the fainter clusters might be
reddened young objects masquerading as old GCs. Further data will be
needed before a clear picture emerges.

Other important clues to the role of disruption processes in shaping
the GCLF may come from the populations of faint extended star clusters
recently discovered in several S0-type galaxies 
\citep{lb2000,larsen2001,peng2006}. These clusters have disk-like 
orbits \citep[at least in one case;][]{bl2002} and spatial distributions 
and effective radii
of 7--15 pc, much larger than the $\sim3$ pc typical of GCs
\citep{jordan2005}. Interestingly, these objects are mostly \emph{fainter}
than the GCLF turn-over, and their LF does not display a turn-over down to
the detection limit of current studies ($M_V\sim-6$). Being more
extended, the two-body relaxation time will be longer, and disk-like
orbits would reduce the role of disk- and bulge shocks. However,
GMC encounters and spiral arm shocks might still have been important
at earlier stages in the history of the host galaxies.  Thus, these
clusters might provide clues to both dynamical destruction processes
and the evolution of S0-type galaxies.

\subsection{Intermediate-age cluster populations}

Important insight into the dynamical evolution of the MF might be
gained from populations of clusters with intermediate ages
($\sim$several Gyrs) and a small age spread.
Among the best candidates for hosting such populations are remnants of major
spiral-spiral mergers. These are known to produce rich populations of
star clusters, although confusion with ancient GCs remains a
difficulty. The nearest such system is NGC~1316, a $\sim3$ Gyr old
merger remnant, whose cluster population has been studied in
detail by \citet{goud2004} using deep HST/ACS imaging.  The metal-poor GCs 
have a normal GCLF, while Goudfrooij et al. found that the metal-rich clusters
(believed to have been formed in the merger) displayed a LF more consistent 
with a power-law down to the 50\% completeness limit at $M_V\sim-6$.
Only in the central regions did they see some evidence for a flattening
of the GCLF at the faint end for the metal-rich GCs. This would be
consistent with the expectation that dynamical processes should first 
start to erode the MF near the centre, but also raises 
the problem that a radial gradient in the GCLF in NGC~1316 would be contrary 
to observations of old GCs and the early disruption scenario of
\citet{vz2003}.  So the implications for understanding the 
evolution of the GCLF are unclear.

Another candidate for hosting an intermediate-age cluster population is
M82.  Based on HST $BVIJH$ imaging, \citet*{deg2003b} derived 
ages of about 1 Gyr for clusters in the ``fossil starburst'' 
region B.  Their derived MF displays
a turn-over at $10^{5.3} \msun$ which they could not explain as a result
of dynamical evolution from a power-law CIMF, thus suggesting that
the CIMF may have been approximately log-normal.  However, since M82 is 
seen nearly edge-on the extinction correction remains a concern, and
another question is whether a starburst would still remain observable as a 
coherent region after $\sim1$ Gyr. HST $U$-band observations are currently 
under way (P.I.\ L.\ Smith), and should help resolve the issue.

In summary, studies of intermediate-age cluster systems, while
potentially promising, have not yet 
solved the problem of whether or not a power-law CIMF
can evolve towards the mass function of old GCs. 

\section{Summary and Outlook}

For several decades it has been clear that a star cluster, even if left in 
isolation, will gradually disrupt as the velocity distribution approaches 
a Maxwellian form and stars in the high-velocity tail escape. This process 
can be greatly accelerated in the presence of external perturbations such 
as the tidal field of the host galaxy, or tidal shocks from encounters with 
bulges, disks, spiral arms and/or GMCs. In the context of this meeting,
it is also worth emphasising that stellar mass loss may play an important
part for the dynamical evolution of star clusters, especially in the
earliest stages.  

There is now excellent agreement between the observed mass- and age 
distributions of open clusters in the Milky Way and theoretical predictions, 
assuming evolution from an initial power-law mass function. It also
appears that we are well under way to understanding differences in 
disruption time-scales in different environments. However, explaining
the mass distribution of ancient globular clusters remains an unsolved
problem, and this is probably where most work lies in the future.
Once again, the effect of stellar mass loss may turn out to be important,
possibly in combination with a more realistic treatment of the
structural properties of star clusters. An additional complication is
that ancient GCs are subject to destruction mechanisms which are
unimportant for star clusters forming today and evolving in disk galaxies 
(bulge, disk shocks), while in contrast the role of other mechanisms which 
may now be important (spiral arm passages, GMC encounters) is difficult 
to assess for GCs.

\acknowledgements 

I am grateful to Yuri Efremov for providing an English translation of the
Ambartsumian (1938) paper.

\end{document}